\documentclass[TWOCOLUMN]{aa}
\usepackage{graphicx}
\usepackage{txfonts}
\begin{document}
   \title{The Galactic Halo density distribution from photometric survey data: results of a pilot study}

   \subtitle{}

  \author{M. Cignoni \inst{1}, V. Ripepi \inst{1}, M. Marconi \inst{1},
  J. M. Alcal\'a \inst{1}, M. Capaccioli  \inst{1} , M. Pannella \inst{2}, R. Silvotti  \inst{1} } \offprints{M. Cignoni}

   \institute{Osservatorio Astronomico Di Capodimonte, Via Moiariello 16,
   80131 Napoli, Italy \and Max-Planck-Institut f\"ur extraterrestrische Physik,
   Giessenbachstrasse, Postfach 1312, D-85741 Garching, Germany}
%DRAFT 15/1/2006

   \date{Received; accepted}

   \abstract{}
% aims heading (mandatory)
{Our goal is to recover the Galactic Halo spatial density by means of field
  stars. To this aim, we apply a new technique to the Capodimonte Deep Field
  (OACDF, Alcal\'a et al. 2004), as a pilot study in view of the VLT Survey
  Telescope (VST) stellar projects. Considering the unique chance to collect
  deep and wide-field photometry with the VST, our method may represent a
  useful tool towards a definitive mapping of the Galactic Halo.}
%methods heading (mandatory)  
{In the framework of synthetic stellar populations, turn-off stars are used to reconstruct the spatial density. The determination of the space density is achieved by comparing the
   data with synthetic color-magnitude diagrams (CMDs). The only assumptions
   involve the IMF, age and metallicity of the synthetic halo
   population. Stars are randomly placed in the solid angle. The contributions
   of the various Monte Carlo distributions (with a step of 4 kpc) along the
   line of sight are simultaneously varied to reproduce the observed CMD.}
% results heading (mandatory)
{Our result on the space density is consistent with a power-law exponent $n\approx 3$ over a range
  of Galactocentric distances from 8 to 40 kpc.}  {}

 \titlerunning{Structure of the Galactic Halo}
\keywords{Galaxy: halo, Galaxy: structure, Galaxy: stellar content, Stars: Hertzsprung-Russell (HR) and C-M, Methods: statistical} 

\maketitle
%
%________________________________________________________________
\section{Introduction}
The astrophysical interest for the spatial distribution of the Galactic halo
is fueled by many important questions concerning the origin of our Galaxy and
its cosmological implications. In the classical scenario (Eggen, Lynden-Bell
\& Sandage \cite{egg}) the origin of the halo is identified in a rapid collapse of
a proto-galactic cloud. An alternative scenario (Searle \& Zinn \cite{sea})
suggests a different history: the Galaxy (and the halo) was built from smaller
blocks, where the stars have already started developing. Signs of the
accretion process are often identified in the observed streams of gas and
stars, remnants of tidally-disrupted satellite galaxies. The emerging picture,
suggested by recent results, has an hybrid nature: the outer halo may have
been formed from the accretion of satellite galaxies, while the inner halo is
thought to be a relic of a dissipative collapse. An evidence in favour of this
scenario is that the axis ratio seems to vary
from inner to outer regions (see e.g. Sommer-Larsen \& Zhen \cite{som}, Gould, Flynn
\& Bahcall \cite{gou}). Variations of the $[\alpha/Fe]$ ratio abundance are also
observed: stars with lower $[\alpha/Fe]$ tend to stay at larger
orbits than the richer ones, suggesting that the outer parts of the halo may
have been accreted by intruder galaxies (see e.g. Nissen \& Schuster \cite{nis}).

In this context, the observation of a particular decline in the density slope
could help to identify the right scenario. The mass density profile has been
extensively studied, both using N-body simulations and matching suitable
stellar tracers. Different power law indexes $n$ ($r^{-n}$) are found:
  
 \begin{itemize}

 \item $n\approx 2$ is required to explain the flat rotation curve;
 \item $n$ between 2.5 to 3.8 (e.g. Avila-Reese et al. \cite{avi}) is supported by N-body simulations for outer halos;
 \item $n\approx 3$ is found from satellite galaxies (SDSS data)
 orbiting around isolated galaxies (Prada et al. \cite{pra});
   
 \item $n\approx 3.3$ is found from local halo stars (Sommer-Larsen \& Zhen \cite{som}) in
   the galactocentric interval 8-20 kpc;
 \item $n\approx 3$ is derived from RR Lyrae stars (Vivas \& Zinn \cite{viv},
 Ivezic et al. \cite{ive}) in the galactocentric range 4-60 kpc;
 \item $n\approx 2.7$ (Siegel et al. \cite{sie}) and $n\approx 2.5 $ (Robin et
 al. \cite{rob}) from star counts.
   \end{itemize}

These estimates suggest that different ``candles'' may yield different
results. In particular it is hard to reconcile the isothermal profile with the
steeper declines derived both from simulations and from stellar tracers. The
goal of this paper is to extend these works, including also turn-off stars. 
In particular, we propose to recover the halo spatial density by comparing the observed CMD
with a synthetic halo population (Monte Carlo generated). Our result is a pure
density distribution of stars along the line of sight (no analitycal a
priori halo density is adopted). Only a posteriori fitting is performed on
the recovered density.
  
Section 2 is dedicated to the sample selection. In
Section 3 we discuss the methodology. In section 4 the reconstruction of the
space density is
fitted with a power law and the robustness of the result is tested against
different assumptions on the adopted IMF and metallicity. Also the possibility
of a metallicity gradient is considered. The conclusions close the paper.

\section{Data selection}
The OACDF is a multi-color imaging survey mainly for extragalactic studies, acquired with the Wide Field Imager (WFI) at the ESO 2.2-m telescope. This
field is located at high galactic latitude ($l\sim 250,\,b\sim 50$), making available a
collection of halo stars with low contamination of disk and thick disk
stars. The field of view covers an area of about $0.5\, \deg ^2$ in the {\it
  B, V, R} optical bands, reaching the following limiting magnitudes: $B_{AB}\sim 25.3$,
$V_{AB}\sim 24.8$ and $R_{AB}\sim25.1$. The completeness magnitude limit is
found at $V_{AB}\sim 24.0$ (for a S/N ratio 10).
 
\subsection{Star/galaxy separation}
\begin{figure}
\includegraphics[width=7 cm]{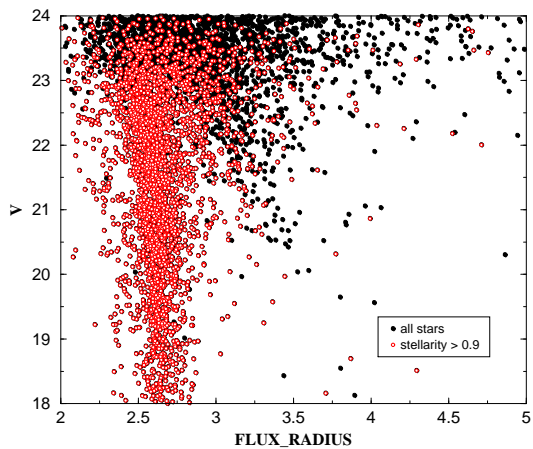}
\caption{Visual apparent magnitude versus the sextractor parameter
  \emph{FLUX\_RADIUS} (two different \emph{CLASS\_STAR} selections are
  shown: filled circles are all objects, open circles are objects with stellarity larger than 0.9).}
\label{1} 
\end{figure}

Contamination by galaxies dominates counts at faint magnitudes. Thus, adequate
star/galaxy separation is essential to infer the Galactic structure. In order
to remove the bulk of extra-galactic contaminants, we applied a filtering based on
the SExtractor (Bertin \& Arnouts \cite{ber}) stellarity parameters (\emph{CLASS\_STAR}) and half-light radius (\emph{FLUX\_RADIUS}). The
point-like sources are identified by choosing objects with stellarity larger
than $0.9$ (simultaneously in {\it B} and {\it V} filters). Beyond the limit $V=23$,
the photometric error in the {\it V} filter (see Alcal\'a et al. \cite{alc}) overcomes
0.05 mag, making difficult a clean separation. This finding is evident in
Figure \ref{1}, where the stellarity selection is performed in a
diagram {\it V} magnitude versus \emph{FLUX\_RADIUS}: selecting stars with high
stellarity removes objects with high radius (candidate galaxies). A side effect of
this procedure is a loss of very faint stars ($V>23$). Therefore, we
limit our analysis to objects brighter than $V=23$. 

The prescription on the stellarity (plus the cut in magnitude) does not ensure
the complete removal of all no-stellar objects. Some point-like sources like
quasars may be still present. To solve this problem we perform a cut in {$
B-V$} (rejecting all objects bluer than 0.3) with a minimal effect on the
stellar population (about $60$ objects removed) \footnote{Halo stars are old and metal poor: assuming the
typical {$B-V$} color of old globular cluster, we don't expect to find stars
bluer than $B-V\approx 0.3-0.4$ (except a negligible fraction of white dwarfs
and HB stars).}.

\subsection{Choice of a stellar tracer}
In order to obtain the spatial density of the Galactic halo one must identify
stars in a precise evolutionary phase and use them to trace the observed CMD
density. The strong requirement on the tracer used to reconstruct the space
distribution is that it must be univocally linked to the distance, once metallicity, mass and age are fixed.
In this context, red giants have an important negative effect, because of the
degeneracy (in color) with main sequence stars. In contrast, turn-off stars have many advantages: 
\begin{itemize}
\item They can be univocally selected (definitely bluer than halo red giants); 
\item They are readily distinguishable from the disk stars in the main sequence (only
  very young and massive disk stars may have similar colors, but they are rare);
\item They are numerous, in fact they outnumber HB stars and red giants by a
  factor $\gtrsim 100$. 
\end{itemize}

\begin{figure}
\centering
\includegraphics[width=7 cm]{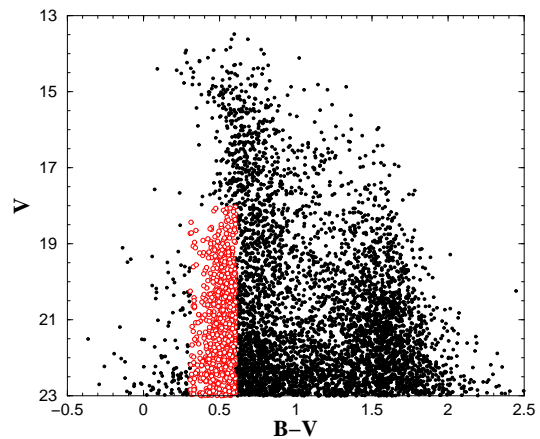}
\caption{Color-magnitude diagram of the point-like sources in the OACDF. The
  turn-off stars discussed in the text are represented with open symbols. } 
\label{to} 
\end{figure}

On this basis, we decided to use turn-off stars as tracers of space
distribution. In order to select turn-off stars we define all object in our
sample with {$B-V$} color in the range 0.3-0.6 (see figure \ref{to}). The
  final data contains about 600 stars.

\subsection{Explored distances}
In what follows we present results for stars in the range $18<V<23$. This
achieves two goals: the bright limit was set to avoid thick disk stars, while
the faint limit minimizes the contamination of no-stellar sources. Moreover,
correlations due to photometric errors are also reduced. These boundaries guarantee a
theoretical range between 4 kpc and 60 kpc \footnote{A simple population with metallicity Z=0.001 and 12 Gyr, the turn-off luminosity
is near to $M_V\sim 4$ (lower metallicities lead to a brighter limit). In this
case, a minimum luminosity $V=23$ implies a maximum distance around 60 kpc.} (heliocentric). 

\section{The model}
A new method for calculating and displaying patterns of Galactic halo stellar
density is presented. The recovered distribution $N (r_{\odot})$ is the number
of stars for heliocentric interval $\delta r_{\odot}$ and solid angle
$\omega$. To recover the spatial density behind the observational CMD we need
to model a synthetic halo population with various theoretical
ingredients. Then the match between theoretical and observational CMDs is
quantified (see
Cignoni et al. \cite{cig} for details).
 
Following a Monte Carlo technique, the theoretical CMD is populated with a
large number of stars randomly built by means of a set of theoretical ingredients. Ignoring chemical evolution, the
synthetic halo population is built near coeval and mono metallic. The adopted
star formation rate is constant between 10 and 12 Gyr (age estimations for
globular clusters indicate values between 11 Gyr, Chaboyer et al. \cite{cha} and 13 Gyr, Hansen et al. \cite{han}). The metallicity is fixed at $[Fe/H]\sim
-1.6$  (see e.g. Ryan \& Norris \cite{rya}) with a Gaussian spread
($\sigma_{[Fe/H]}=1.0$). The adopted IMF is a power law
with a Salpeter index. Once masses, ages and
metallicity are known, colors and magnitudes are obtained by interpolating a
library of stellar evolutionary tracks (Pisa Evolutionary Library, Cariulo et
al. \cite{car}). 

In order to reconstruct the halo structure, this synthetic population is
placed at different distances, randomly placing stars in the $j$-th heliocentric
interval $[d_{\odot ,i},\,d_{\odot, i}+4\,kpc]$. The synthetic diagrams are
corrected for reddening and extinction according to Schlegel et al. \cite{sch} maps
(for the OACDF field, a $E(B-V)\approx 0.05$ is adopted). The final product
is a base of partial color magnitude diagrams, each one representing the same
population but at different distance moduli.
Transforming these CMDs in two dimensional histograms (bin size: $\delta V=0.05 $, $\mathrm{mag},
\delta (B-V)=0.05\, \mathrm{mag}$), we obtain a base in a 2-D space ({\it V}
versus {$B-V$}) and we can write every CMD as a linear
combination of these histograms. In numbers, a generic CMD is so built:

\begin{eqnarray}  
&&m_{i}=\sum_{j} \alpha_{j}\,c_{ij}\label{rl3}
\end{eqnarray} 

where $m_{i}$ is the number of star in the final CMD in bin $i$, $\alpha_{j}$ is
the stellar rate in the distance slab $j$, and $c_{ij}$ is the number of
stars in the bin $i$ owning to the partial CMD $j$.

The combination of coefficients $\alpha_{j}$ that minimize a Poisson based
likelihood is searched through a simplex algorithm. 
%\begin{table}
%\centering
%\begin{tabular}{ll}
%\hline \hline\noalign{\smallskip}
%        & \large $L= \sum_{i=1}^{N bin} O_{i}\ln\frac{O_{i}}{M_{i}}+O_{i}-M_{i}$\\
%        & $O_{i}$=observed objects in the i-bin\\
%        & $M_{i}$=theoretical objects in the i-bin\\
%   \hline
%    \hline
%\end{tabular}{\smallskip}{\smallskip}{\smallskip}{\smallskip}\caption{Likelihood function.}
%\label{tab}
%\end{table}
The uncertainties on the final coefficients are found by using a bootstrap
technique. The final result is the distribution of stars along the line of
sight in the direction of the OACDF. However, real data are observed in solid
angle, therefore our result is not a density yet. After correcting for the
solid angle, the space density as a function of the Galactocentric distance is
found by converting from heliocentric distances to Galactocentric distances.

\section{Results}

\begin{figure}
\centering
\includegraphics[width=6.9cm]{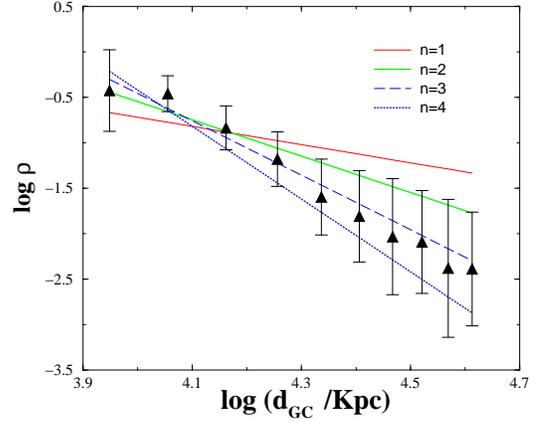}
\caption{The recovered density distribution (the density scale is somewhat
  arbitrary). Error bars are $1\,\sigma$. For comparison, different power law densities are also shown
  (with the labeled exponents).}
\label{4} 
\end{figure}

The recovered spatial distribution is shown in Figure \ref{4} (symbols with
error bars). We have to
stress the exact nature of our result. In fact, the OACDF field \emph{is not
along a galactic radius}, but it is rather oblique. Thus, comparing our result
with simple radial profiles means that \emph{we are assuming an axially
  symmetric Galaxy}. Moreover, due to the large uncertainties of the
further bins, our analysis is truncated at the galactocentric distance of 40 kpc.

A parametrized model of the mass distribution is fitted to the recovered
density. The halo is described by a power law $r^{-n}$ ($r$ galactocentric
radius), with axis ratio equals to 1 (round halo). 

Different hypothesis on the power law index $n$ (see figure \ref{4}) are
explored through a chi square fitting technique.

\begin{figure}
\centering
\includegraphics[width=6.9cm]{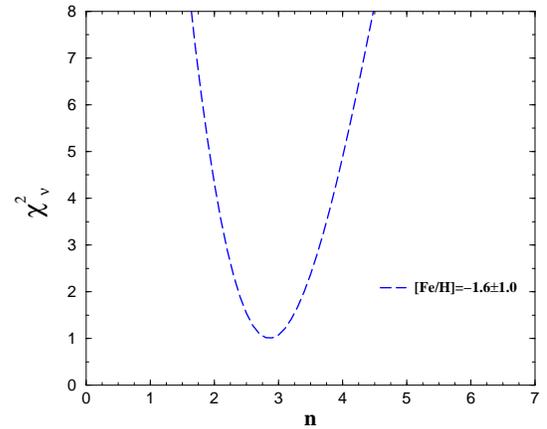}
\caption{Reduced $\chi^2$ versus the {\it n} exponent.} 
\label{5} 
\end{figure}

Figure \ref{5} illustrates the reduced $\chi^2_\nu$ ($\chi^2$/degrees of freedom) versus the power law
index $n$: the minimum is placed near $n=2.9$ with $\chi^2_\nu\approx 1.0$.  
Although it is difficult to define a confidence interval\footnote{The
uncertainties on the recovered densities may hide correlations among bins
(incompleteness, external contaminations and wrong metallicitiy may produce
systematic effects; see e.g. Cignoni et al. \cite{cig}), making the $\chi^2$
probability quite useless.} for $n$, these results confirm a stellar halo with
a decline \emph{steeper than an isothermal profile} (which is the accepted
profile for the dark halo). Indeed, values $n$ lower than $2$ give a
$\chi^2_\nu$ more than $4$ times higher than the $n=3$ result.

\subsection{Alternative scenarios}
In order to avoid biases due to the assumed halo properties, namely the IMF, the metallicity \footnote{In principle also the assumed age is
  critical. However, the small masses involved (lower than $0.8\, M_{\odot}$)
  imply a weak sensitivity to the age.} and the axis ratio, we have repeated our recovery by
  using different scenarios.
\subsection{IMFs}
\begin{figure}
\centering
\includegraphics[width=6.9cm]{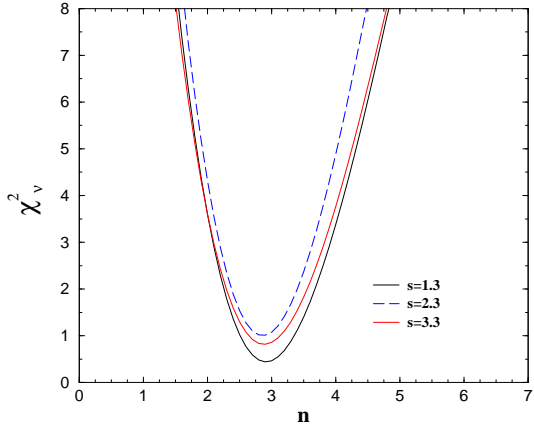}
\caption{Reduced $\chi^2$ versus index {\it n} for different assumed IMF (the
  explored exponents are labeled).} 
\label{6} 
\end{figure}

Figure \ref{6} shows the effect of different IMFs.  
The result is quite independent of the precise IMF ($dN/dm \propto m^{-s}$) exponent $s$, probably due to
the short mass spectrum involved in the halo population. The cuts at $V=23$ and
$B-V=0.6$, combined with the typical low metallicities involved in the halo,
imply a detectable minimum mass of about $0.6\,M_{\odot}$. Then, the age of
the halo, that is higher than 10 Gyr, implies a maximum mass (for stars still in main
sequence) around $0.8\,M_{\odot}$. In conclusion, the total mass spectrum is
only $\approx 0.2$ solar masses: too short to distinguish between different IMF
slopes. Photometric errors, the presence of different metallicities and low
number statistics can largely sweep out an IMF effect.

\subsection{Metallicity effect}

%The effect of a different metallicity is explored (see figure \ref{7}).
\begin{figure}
\centering
\includegraphics[width=6.9cm]{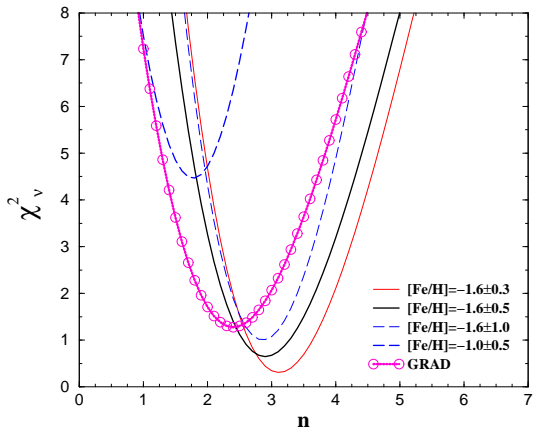}
\caption{Reduced $\chi^2$ versus index {\it n} for several assumed halo
  metallicities. The
  metallicity distribution is always a Gaussian with the labeled mean and
  dispersion; the label ``grad'' indicates a model with a spatial gradient - see text.}
\label{7} 
\end{figure}
A variation in the metallicity spread has not a significant impact (figure \ref{7} shows the result for $\Delta
[Fe/H]=0.3,0.5,1.0$). In fact, a metallicity spread introduces a blurring on
the partial CMDs and this causes an additional random uncertainty on the
recovered density. However, this effect is not systematic.

The result is different when the mean metallicity of the model is changed. In
particular, increasing the metallicity to $[Fe/H]=-1.0$ shifts the most probable $n$ exponent
to lower values. The new best value is close to $n=2$, but the relative
$\chi^2_\nu$ is substantially higher than the previous estimates ($\approx
4.5$).

Finally, Figure \ref{7} shows also the resulting $\chi^2_\nu$ when an extreme
spatial gradient (0.2 dex each 4 kpc, $[Fe/H]=0$ in the Solar position) is
adopted (labeled with ``grad''). As expected, the adoption of a metallicity
gradient has an intermediate effect: the minimum $\chi^2_\nu$ is reached for
$n\approx 2.3$, yielding a $\chi^2_\nu$ value around 1.3, which is comparable
with those resulting from the first models.
\subsection{Axis ratio}
\begin{figure}
\centering
\includegraphics[width=6.9cm]{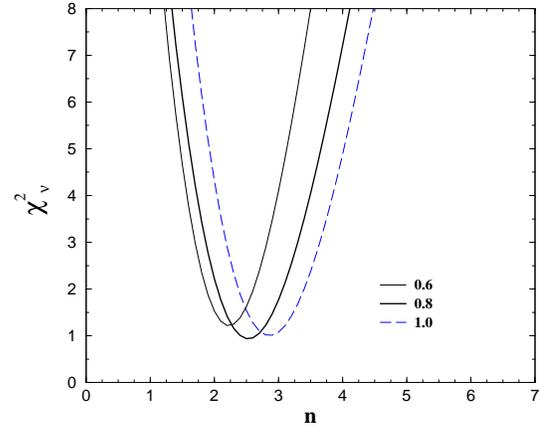}
\caption{Reduced $\chi^2$ versus index {\it n} for several hypothesis on the
  axis ratio (AR= 0.6, 0.8, 1.0).}
\label{axis} 
\end{figure}

In this section the hypothesis of a round halo is abondoned. Although the
axis ratio (AR) is not significantly constrained by a single field, two
alternative values are explored (AR = 0.8 and 0.6). Figure \ref{axis} shows
the results. Decreasing the AR determines a lower best value for
$n$. The extreme choice AR=0.6 implies a value $n\approx 2.2$. It is
noteworthy that the best $n$, independently of the precise AR, is caracterized
by similar chi square values: for a single field, the degeneracy between axis
ratio and power law index does not allow to disentangle the two effects.

\section{Conclusions}

We have combined a star count model with an updated stellar library to
reconstruct the Galactic Halo stellar density in the direction of the
Capodimonte Deep Field. Our results confirm a steep decline. In particular,
when a metallicity $[Fe/H]=-1.6\pm1.0$ is adopted, a power-law index $n\approx
3$ gives a reasonable fit out to $\approx 40$ kpc from the galactic
centre. This is quite different from the value $n\approx 2$, implied by the
flat rotation curve. However, gas and turn off stars may trace different halo
regions. First, the measurements of rotation velocity from gas are reliable up
to about 20 kpc from the Galactic Center while turn off stars reach 40 kpc,
suggesting that our exploration is deeper in the outer halo. Second, the OACDF
field is high on the Galactic plane while the gas is locked on the plane.

The value $n\approx 3$ is also larger than the Robin et al. \cite{rob} and
 Siegel et al. \cite{sie} findings, that analyzed several galactic fields
 (generally shallower than OACDF) using similar stellar tracers. Basically,
 our study confirms the results obtained with RR Lyrae stars.

An higher metallicity ($[Fe/H]=-1.0$) can lower our best value to the
isothermal $n\approx 2$, but the corresponding $\chi^2_\nu$ is significantly
worse. Moreover, the presence of an extreme chemical gradient ($\chi^2_\nu$
values similar to the $[Fe/H]=-1.6\pm1.0$ scenario) cannot be ruled out.

A lower index $n$ is also obtainable by adopting smaller axis ratios. In
  these cases, the minimum $\chi^2_\nu$ are close to case with round
  halo (AR=1.0), leaving the possibility of a lower index $n$ ($n=2.2$ for AR=0.6).

This work represents a pilot program for a larger
project devoted to the exploration of the VST halo fields (STREGA@VST, see
Marconi et al. \cite{marco}). The possibility to perform multi-direction comparisons
will allow us to reconstruct the 3-D pattern of the halo and to disentangle
the degeneracy between the axis ratio and the density power law slope. Moreover, the
availability of several photometric filters will provide a chance to improve
both the comparison between synthetic and observational CMDs and the
separation star/galaxies.

\begin{acknowledgements}
We are greatful to Steven N. Shore for his interest in and support of this
work, and to S. Degl'Innocenti and P. G. Prada Moroni for helpful discussions
and for kindly providing us specific evolutionary tracks. We thank the OACDF
team (http://www.oacn.inaf.it/oacdf-bin/cdfcgi?people) for their contribution
on data reduction and discussions. Financial support for this study was
provided by MIUR, under the scientific project ``On the evolution of stellar
systems: fundamental step toward the scientific exploitation of VST''
(P.I. Massimo Capaccioli).
\end{acknowledgements}


\begin{thebibliography}{}
\bibitem[2004]{alc}Alcal\'a, J. M., Pannella, M., Puddu, E., et al. 2004, A\&A, 428, 339
\bibitem[1999]{avi}Avila-Reese V., Firmani C., Klypin A., Kravtsov A. V.,
  1999, MNRAS, 310, 527
\bibitem[1996]{ber} Bertin, E., \& Arnouts, S. 1996, A\&AS, 117, 393 
 \bibitem[2004]{car}Cariulo, P., Degl'Innocenti, S., \& Castellani, V. 2004,
 A\&A, 421, 1121
\bibitem[2006]{cig}Cignoni, M., Degl'Innocenti, S., Prada Moroni, P. G.,
  Shore, S. N., 2006, A\&A, in press
\bibitem[1998]{cha}Chaboyer, Brian, Demarque, P., Kernan, Peter J., Krauss,
  Lawrence M.,1998, ApJ, 494, 96

\bibitem[1962]{egg}Eggen, O. J., Lynden-Bell, D. \& Sandage, A. R. 1962, ApJ,
  136, 748
\bibitem[1998]{gou} Gould, A., Flynn, C. \& Bahcall, J. N. 1998, ApJ, 503, 798
\bibitem[2002]{han}Hansen, Brad M. S. et al. 2002, ApJ, 574, 155
\bibitem[2004]{ive}Ivezic et al. 2004, ASP Conf. Ser. 327, Satellite and Tidal
  Streams, ed. F. Prada, D. Martinez-Delgado, \& T. Mahoney (San Francisco:
  ASP), 104
\bibitem[2006]{marco}Marconi M. et al., 2006, Mem. della Soc. Astron. Ital. Supp., 9, 253
\bibitem[1997]{nis}Nissen P. E., Schuster W. J., 1997, A\&A, 326, 751
\bibitem[2003]{pra}Prada, F., et al. 2003, ApJ, 598, 260
\bibitem[2000]{rob}Robin A. C., Reyl\'e C., Cr\'ez\'e M., 2000, A\&A, 359, 103
\bibitem[1998]{sch}Schlegel, D. J., Finkbeiner, D. P., \& Davis, M. 1998, ApJ, 500, 525
\bibitem[1978]{sea}Searle, L. \& Zinn, R. 1978, ApJ, 225, 357
\bibitem[2002]{sie}Siegel, M. H., Majewski, S. R., Reid, I. N., Thompson,
  I. B. 2002, ApJ, 578, 151
\bibitem[1991]{rya}Ryan, S. \& Norris, J., 1991, AJ, 101, 1865
\bibitem[1990]{som}Sommer-Larsen, J. \& Zhen, C. 1990, MNRAS, 242, 10
\bibitem[2003]{viv}Vivas A. K., Zinn R., 2003, Mem. Soc. Astron. Ital., 74, 928
\end{thebibliography}
\end{document}